%% file: main.tex
\documentclass[conference]{IEEEtran}
\IEEEoverridecommandlockouts

\usepackage{cite}
\usepackage{amsmath,amssymb,amsfonts}
\usepackage{algorithm}
\usepackage{algpseudocode}
\usepackage{graphicx}
\usepackage{textcomp}
\usepackage{enumitem}
\usepackage{xcolor}
\usepackage{makecell}
\def\BibTeX{{\rm B\kern-.05em{\sc i\kern-.025em b}\kern-.08em
    T\kern-.1667em\lower.7ex\hbox{E}\kern-.125emX}}
\begin{document}

\title{GPU Acceleration of TFHE-Based High-Precision Nonlinear Layers for Encrypted LLM Inference}


\author{
\vspace{-2pt}
    {\emph{Guoci Chen$^{\ast}$ \hspace{8pt} Xiurui Pan$^{\ast}$ \hspace{8pt} Qiao Li$^{\dagger}$ \hspace{8pt} Bo Mao$^{\ddagger}$}}\\
\vspace{-1pt}
    {\emph{Congming Gao$^{\ddagger}$ \hspace{8pt} Chengying Huan$^{\sharp}$ \hspace{8pt} Mingzhe Zhang$^{\diamondsuit}$ \hspace{8pt} Jie Zhang$^{\ast}$}}\\
\vspace{-1pt}
    {Computer Hardware and System Evolution Laboratory}\\
\vspace{-1pt}
    {\emph{Peking University$^{\ast}$, Mohamed bin Zayed University of Artificial Intelligence$^{\dagger}$}}\\
\vspace{-1pt}
   {\emph{Xiamen University$^{\ddagger}$, Nanjing University$^{\sharp}$, Ant Group$^{\diamondsuit}$}}\\
\vspace{-1pt}
   {\emph{https://www.chaselab.wiki}}\\
}

\maketitle

\begin{abstract}

Deploying large language models (LLMs) as cloud services raises privacy concerns as inference may leak sensitive data. Fully Homomorphic Encryption (FHE) allows computation on encrypted data, but current FHE methods struggle with efficient and precise nonlinear function evaluation. Specifically, CKKS-based approaches require high-degree polynomial approximations, which are costly when target precision increases. Alternatively, TFHE's Programmable Bootstrapping (PBS) outperforms CKKS by offering exact lookup-table evaluation. But it lacks high-precision implementations of LLM nonlinear layers and underutilizes GPU resources.

We propose \emph{TIGER}, the first GPU-accelerated framework for high-precision TFHE-based nonlinear LLM layer evaluation. TIGER offers: (1) GPU-optimized WoP-PBS method combined with numerical algorithms to surpass native lookup-table precision limits on nonlinear functions; (2) high-precision and efficient implementations of key nonlinear layers, enabling practical encrypted inference; (3) batch-driven design exploiting inter-input parallelism to boost GPU efficiency. TIGER achieves 7.17$\times$, 16.68$\times$, and 17.05$\times$ speedups over a CPU baseline for GELU, Softmax, and LayerNorm, respectively.

\end{abstract}

\section{Introduction}
\input{intro.tex}

\section{Background}
\input{background.tex}

\section{Motivation}
\input{motivation.tex}

\section{Design}
\input{design.tex}

\section{Evaluation}
\input{evaluation.tex}

\section{Discussion}
\input{discussion.tex}

\section{Conclusion}
\input{conclusion.tex}

\bibliographystyle{IEEEtran}
\bibliography{refs}

\end{document}

%% file: intro.tex
\label{sec:intro}

Large language models (LLMs) have emerged as key components of various applications. Thanks to their popularity, LLMs are widely deployed as cloud-hosted services where users send their data and queries (usually in plaintext) to remote servers for processing. This deployment strategy raises significant privacy concerns, as sensitive user data may be exposed to the service provider. Fully Homomorphic Encryption (FHE) schemes, such as CKKS~\cite{ckks} and TFHE~\cite{tfhe}, offer a promising solution by enabling computation directly on encrypted data. By doing so, servers can perform inference without ever accessing the plaintext inputs.

Recent works have explored encrypted LLM inference based on CKKS, leveraging its efficient support for approximate arithmetic and SIMD-style batching to accelerate linear layers (e.g., matrix multiplications in attention and feed-forward networks)~\cite{encrypted-llm}. Unfortunately, CKKS does not natively support nonlinear functions, which are instead approximated using high-degree polynomials. This introduces a fundamental trade-off: achieving high numerical fidelity requires increasing the polynomial degree, which, in turn, inflates multiplicative depth, bootstrapping frequency, and key sizes~\cite{lee2022ckksboot, autofhe}. As a result, high-precision nonlinear evaluation becomes prohibitively expensive in both computation and memory, making it the primary bottleneck in CKKS-based encrypted LLM inference~\cite{autofhe}.

An alternative approach is to leverage the TFHE scheme, which evaluates nonlinear functions via Programmable Bootstrapping (PBS). PBS enables exact lookup-table (LUT) evaluation over encrypted data, avoiding polynomial approximation entirely. To support higher precision, prior work proposes advanced techniques such as Without-Padding PBS (WoP-PBS) and FBT-TFHE~\cite{fbt-tfhe}. Recent implementations of TFHE-based libraries (e.g., TFHE-rs~\cite{TFHE-rs}) have provided primitive-level operators (e.g., KeySwitch and PBS) and basic integer arithmetic operations (e.g., addition and multiplication), which enable the construction of more complex functions.

Despite these advances, existing TFHE-based libraries remain insufficient for practical encrypted LLM inference.
Specifically, realizing high-precision nonlinear evaluation under TFHE poses three key challenges:

\noindent $\bullet$ \emph{Challenge 1: Precision Limitation of Current Methods.}
Although advanced LUT-based techniques, such as WoP-PBS, enable higher-precision evaluation, their effective input precision remains limited (typically up to $\sim$20 bits), which constrains the precision in nonlinear function evaluation in LLM nonlinear layers.

\noindent $\bullet$ \emph{Challenge 2: Lack of High-Level Layer Solutions.}
The state-of-the-art TFHE frameworks only offer low-level operators. There exists no concrete practice to build nonlinear functions or their composition into LLM layers, leaving a substantial gap between primitive-level support and model-level deployment.

\noindent $\bullet$ \emph{Challenge 3: Underutilization of Inter-Input Parallelism. }
Prior work mainly exploits intra-operation parallelism within a single group of ciphertexts, while overlooking the large degree of inter-input parallelism across independent activations in LLM layers. This results in severe underutilization of GPU parallelism, e.g., unbatched GELU layer evaluation is 4.35$\times$ slower than the batched version.

To address these challenges, we propose \emph{TIGER}, the world-first GPU-accelerated framework for TFHE-based high-precision evaluation of LLM nonlinear layers. 
TIGER bridges the gap between cryptographic primitives and model-level execution by co-designing high-precision LUT evaluation, numerical methods, and GPU parallelization.

At the core of TIGER is a GPU implementation of the WoP-PBS algorithm, which enables efficient high-precision LUT evaluation. Building upon this primitive, we further integrate numerical techniques to extend the precision of function evaluation beyond the native LUT resolution, achieving higher accuracy with minimal additional overhead. Using these building blocks, TIGER constructs a set of key nonlinear functions, including $\exp(x)$, $\mathrm{GELU}(x)$, and $1/\sqrt{x}$, and further composes them into high-level LLM layers such as Softmax, LayerNorm, and GELU layers.

To fully exploit GPU computational parallelism, TIGER introduces a batch-oriented execution model that processes a large number of independent ciphertext inputs simultaneously. By carefully organizing ciphertext groups and scheduling PBS operations in batches, TIGER exposes inter-input parallelism across LLM layers, significantly improving hardware utilization. In addition, we apply a series of GPU-specific optimizations to further enhance throughput and reduce execution overhead.

Compared to optimized CPU-based implementations, TIGER achieves substantial performance improvements, delivering $7.17\times$, $16.68\times$, and $17.05\times$ speedups for GELU, Softmax, and LayerNorm, respectively. For end-to-end evaluation of a GPT-2 Transformer block, TIGER achieves up to $15.54\times$ speedup.

Our contributions are summarized as follows:

\begin{itemize}[leftmargin=10pt]
\setlength{\itemsep}{5pt}

\item \emph{World-first TFHE-based high-precision nonlinear framework for LLMs.}
We present the world-first systematic framework for evaluating LLM nonlinear layers (Softmax, LayerNorm, GELU for GPT-2 model) with high precision under the TFHE scheme, bridging the gap between low-level primitives and model-level execution.

\item \emph{High-precision nonlinear function evaluation via LUT-numerical co-design.}
We combine high-precision LUT evaluation with numerical methods to implement key nonlinear functions (e.g., $\exp$, $\mathrm{GELU}$, and inverse square root) with improved accuracy and efficiency. By synergistically integrating lookup table techniques with advanced numerical algorithms such as range reduction and iterative refinement, we achieve precision levels that significantly exceed the native LUT resolution while maintaining practical computational overhead. This co-design approach enables accurate evaluation of complex activation functions required by modern transformer architectures.

\item \emph{GPU-native high-precision TFHE computation substrate.}
We construct the WoP-PBS algorithm on GPU and develop a set of fixed-point arithmetic primitives to support high-precision encrypted computation, forming an efficient foundation for nonlinear function evaluation. Our GPU-based solution leverages its massive parallel processing capabilities to accelerate the computationally intensive bootstrapping operations. We further customize a fixed-point arithmetic library to provide the essential building blocks for constructing sophisticated encrypted computations that were previously impractical on resource-constrained platforms.

\item \emph{Hardware-aware optimization for TFHE-based nonlinear evaluation.}
We design and implement a set of hardware-aware optimizations that adapt TFHE computation to GPU architectures, improving parallel efficiency, resource utilization, and overall system throughput. Our optimizations include intelligent batch scheduling to maximize GPU occupancy, memory coalescing strategies for efficient data movement, and warp-level primitives that exploit the underlying hardware parallelism. These optimizations collectively reduce latency and make encrypted LLM inference feasible at practical performance levels.

\end{itemize}

%% file: background.tex
\label{sec:background}

\subsection{FHE Basics}
\label{sec:bg:fhe}

Fully Homomorphic Encryption (FHE) enables arbitrary computation over encrypted data without revealing the exact plaintext. Among existing FHE schemes, \emph{CKKS}~\cite{ckks} and \emph{TFHE}~\cite{tfhe} represent two dominant paradigms with distinct design trade-offs.

\noindent \textbf{CKKS.}
The CKKS scheme~\cite{ckks} supports approximate arithmetic over real numbers and is particularly efficient for linear operations such as additions and multiplications. Its SIMD-style packing allows multiple values to be processed in parallel within a single ciphertext, making it well-suited for linear layers in neural networks. However, CKKS does not natively support nonlinear functions. Instead, it relies on polynomial approximations to evaluate nonlinear functions, which introduces a trade-off between approximation accuracy, multiplicative depth, and bootstrapping overhead.

\noindent \textbf{TFHE.}
In contrast, TFHE~\cite{tfhe} operates over discrete message spaces and enables efficient evaluation of arbitrary functions via Programmable Bootstrapping (PBS). A PBS operation refreshes ciphertext noise while simultaneously applying a lookup-table (LUT) function to the encrypted input. This allows TFHE to evaluate nonlinear functions exactly without resorting to polynomial approximation.

A typical PBS operation involves several key primitives: (1) \emph{Key Switching} (KS), which transforms the input LWE ciphertext to one encrypted under the bootstrapping key; (2) \emph{Blind Rotation} (BR), which rotates a test polynomial by the encrypted message, effectively performing the LUT evaluation in the exponent; and (3) \emph{Sample Extraction} (SE), which extracts the result as a fresh LWE ciphertext from the rotated RLWE ciphertext. Together, these steps simultaneously reduce noise and evaluate an arbitrary function encoded in the LUT.

While TFHE provides a direct mechanism for nonlinear evaluation, its efficiency critically depends on how high-precision LUTs are implemented, which we will discuss shortly.

\subsection{High-Precision LUT Evaluation in TFHE}
\label{sec:bg:lut-eval}

Standard PBS in TFHE supports only small message spaces (typically a few bits), which is insufficient for high-precision nonlinear functions required in LLM inference. To address this limitation, prior work has proposed advanced techniques for high-precision LUT evaluation.

\noindent \textbf{FBT-TFHE.}
Functional Bootstrapping (FBT)~\cite{fbt-tfhe} extends PBS by organizing LUTs in a tree structure, where each level of the tree produces one intermediate result for each LUT by one PBS operation, respectively, and repacks the results as new LUTs for the next level, until the final output is obtained. This approach enables higher input precision; unfortunately, it suffers from the high cost of rapidly increased computational complexity and bootstrapping overhead, especially when the input precision scales.

\noindent \textbf{WoP-PBS.}
Compared to FBT-TFHE, Without-Padding PBS (WoP-PBS)~\cite{wop-pbs} reduces the computational cost of high-precision evaluation by restructuring the computation into a sequence of fine-grained primitives, including Bit Extraction (BE), Circuit Bootstrapping (CB), and Vertical Packing (VP). This approach replaces heavy PBS operations with more lightweight CMux operations, which are more efficient and scalable for high-precision inputs.

Despite these advances, efficiently implementing these algorithms, particularly on parallel hardware such as GPUs, remains a key challenge. WoP-PBS involves multi-stage operations requiring key switching and bootstrapping across three distinct key spaces, whose computational and memory-access complexity makes implementation considerably demanding. On GPU architectures, especially, effectively harnessing parallelism to accelerate these operations presents a non-trivial challenge. To date, while some libraries like TFHE-rs offer CPU-based implementations, a highly efficient and open-source GPU implementation is still lacking, which has hindered the broader adoption and practical deployment of these algorithms in real-world applications.


\subsection{LLM Architecture Introduction}
\label{sec:bg:llm-architecture}

Modern LLMs are typically based on the Transformer architecture~\cite{transformer}, which consists of stacked layers of attention and feed-forward networks. We take GPT-2~\cite{gpt2} as a representative example.

A GPT-2 model stacks multiple such Transformer blocks sequentially, in which we mainly focus on the structure of a single block in the following. Each Transformer block contains two main components: a multi-head self-attention module and a feed-forward network (FFN). The attention module computes query, key, and value projections followed by a Softmax operation to produce attention weights. The FFN consists of two linear transformations with a nonlinear activation function (e.g., GELU) in between. In addition, LayerNorm is applied to stabilize training and inference.

\begin{figure}[t]
	\centering
	\includegraphics[width=\linewidth]{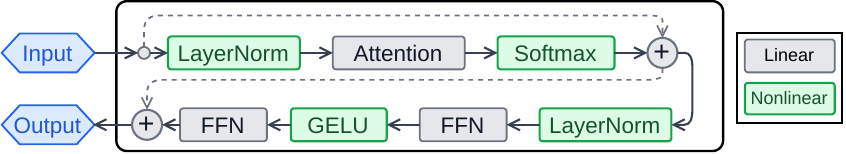}
	\caption{Single GPT-2 Transformer block. Nonlinear components (\emph{LayerNorm}, \emph{Softmax}, and \emph{GELU}) are highlighted, while linear transformations and residual additions are shown in gray.}
	\label{fig:gpt2-architecture}
\end{figure}

Figure~\ref{fig:gpt2-architecture} shows a single GPT-2 Transformer block in a compact layout, with nonlinear layers explicitly highlighted.
From a computational perspective, Transformer inference can be divided into:
(1) \emph{linear operations} (matrix multiplications and additions in attention and FFN layers), and
(2) \emph{nonlinear operations} (Softmax, LayerNorm, GELU).
It is worth noting that nonlinear operations are significantly more sensitive to precision reduction and play a critical role in preserving model accuracy (cf. Section~\ref{sec:motiv:prec-requirement} for details). This makes them the primary target for high-precision evaluation in encrypted inference.

In this work, we aim to accelerate these nonlinear layers under the TFHE scheme by leveraging high-precision LUT evaluation and hardware acceleration.

%% file: motivation.tex
\label{sec:motivation}

\subsection{Precision Requirements of LLM Nonlinear Layers}
\label{sec:motiv:prec-requirement}

Recent advances in model compression have proven that large language models (LLMs) can tolerate aggressive quantization in linear layers while maintaining acceptable accuracy.
For example, prior work such as SmoothQuant~\cite{smoothquant} and GPTQ~\cite{gptq} demonstrates that weights and activations in linear operators can be reduced to low precision (e.g., 4--8 bits) with minimal degradation.
However, these methods consistently treat nonlinear operations (e.g., Softmax, LayerNorm, and GELU) with higher precision or specialized handling, due to their strong numerical sensitivity.
This suggests that nonlinear layers play a critical role in preserving model accuracy during inference.

\begin{table}
\centering
\small
\caption{Impact of nonlinear precision on GPT-2 perplexity (WikiText-2). Lower is better.}
\label{tab:nonlinear-precision}
\begin{tabular}{lrr}
\hline
\textbf{Configuration} & \textbf{PPL} & \textbf{Norm.} \\
\hline
Baseline (20-bit)            & 24.36   & 1.00$\times$ \\ 
\hline
Softmax (4-bit)              & 887.56  & 36.44$\times$ \\
LayerNorm (4-bit)            & 16274.82 & 668.18$\times$ \\
GELU (4-bit)                 & 1401.12 & 57.52$\times$ \\
All nonlinear (naive 4-bit)  & 6322.33 & 259.57$\times$ \\
\hline
PTQ (all 4-bit)              & 2746.04 & 112.74$\times$ \\
QAT (all 4-bit)              & 1267.59 & 52.04$\times$ \\
\hline
Linear-only (NF4)            & 26.33   & 1.08$\times$ \\
Linear-only (FP4)            & 28.89   & 1.19$\times$ \\
\hline
\end{tabular}
\end{table}

To quantify this effect, we conduct an empirical study on GPT-2 by selectively reducing the precision of different nonlinear components and measuring the resulting perplexity (PPL).
Table~\ref{tab:nonlinear-precision} summarizes the results.
We observe that aggressively quantizing nonlinear functions leads to severe accuracy degradation.
Individually quantizing Softmax, LayerNorm, and GELU to 4-bit precision increases PPL by $36.44\times$, $668.18\times$, and $57.52\times$, respectively.
When all nonlinear components are quantized naively, the degradation further amplifies to $259.57\times$.
Even advanced techniques such as PTQ and QAT only partially mitigate the issue, still resulting in over $50\times$ degradation.
In contrast, when only linear layers are quantized while preserving high precision for nonlinear layers, the model maintains near-baseline accuracy, with less than $20\%$ increase in PPL.
This stark contrast highlights that nonlinear layers are significantly more sensitive to precision reduction than linear components.



\noindent \textbf{Takeaway 1:}
\emph{Nonlinear layers in LLM inference require sufficiently high precision to preserve model accuracy, making them the primary bottleneck for low-precision and encrypted evaluation.}

\subsection{Limitation of Existing High-Precision LUT Evaluation Methods}
\label{sec:motiv:lut-eval-limit}

Existing high-precision LUT evaluation methods (e.g., WoP-PBS and FBT-TFHE) offer a precision of $\sim20$ bits for input, and are hard to scale up further due to growing computation costs and ciphertext noise.

\textbf{For FBT-TFHE}, lookup table size and the number of PBS operations in need both grow exponentially with the input precision. We evaluated that given an input of 20-bit precision, 65536 LUT ciphertexts, and 69905 PBS operations are needed for one LUT evaluation operation, which eventually takes $\sim20$s on GPU. Also, the private key switch key needs $\sim10$ GB GPU memory in our configuration, which incurs a large overhead. Given the already significant overhead of FBT-TFHE in 20-bit precision, it's unbearable to elevate the precision further.

\textbf{For WoP-PBS}, LUT evaluation becomes more lightweight, and time and memory are not bottlenecks in this case. However, the variance of noise depends on the size of CMux tree in the vertical packing stage of WoP-PBS, which grows exponentially with the input precision \cite{wop-pbs-2}. If the precision grows further, the noise will disturb the correctness of ciphertexts and result in decryption failure.

\noindent \textbf{Takeaway 2:}
\emph{Existing LUT evaluation methods fail to scale without further algorithm advancement.}

\subsection{Limited Parallelism and Batching of PBS}
\label{sec:motiv:pbs-batch}

Previous integer-level libraries (e.g., TFHE-rs) primarily exploit intra-input parallelism by batching PBS operations within a single input, while overlooking potential inter-input parallelism. Alternatively, packing more PBS operations into a batch appears to increase GPU utilization. Our measurements reveal a fundamentally non-monotonic relationship.

\begin{figure}[t]
    \centering
    \includegraphics[width=\linewidth]{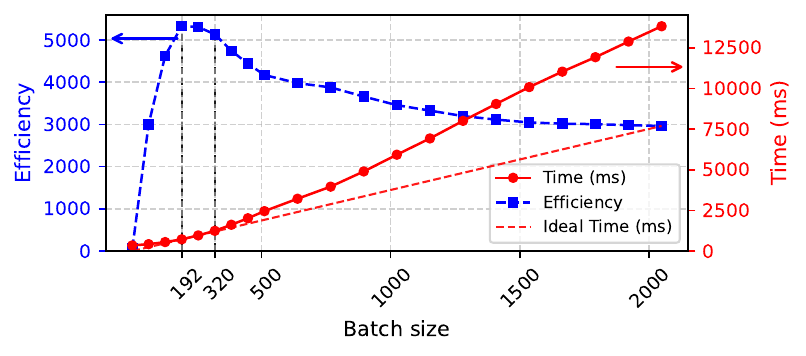}
    \vspace{-15pt}
    \caption{Time (red) and efficiency (blue) of PBS with different batch sizes; ideal linear scaling is shown in dashed red.}
    \label{fig:batching_efficiency}
    \vspace{-10pt}
\end{figure}

We test and collect PBS time and efficiency under different batch sizes ($\text{efficiency} = \text{batch size} / \text{time}$).
Figure~\ref{fig:batching_efficiency} shows that as the batch size increases, efficiency initially improves substantially but begins to decline beyond the optimal range of $[192, 320]$, dropping to 55\% of peak performance at the extremes.
To understand this behavior, we profile PBS kernels with \texttt{ncu}, focusing on the two external multiplication stages. Tables~\ref{tab:small-batch} and~\ref{tab:large-batch} summarize the key metrics.

\begin{table}[t]
\centering
{\footnotesize
\setlength{\tabcolsep}{2.5pt}
\caption{Profiling results for small batch sizes (stage 2 underutilization).}
\label{tab:small-batch}
\begin{tabular}{c|ccc|ccc}
\hline
 & \multicolumn{3}{c|}{\textbf{Overall}} & \multicolumn{3}{c}{\textbf{Stage 2}} \\
\textbf{BS} & \textbf{AW (\%)} & \textbf{IA (\%)} & \textbf{SNS (\%)} & \textbf{AW (\%)} & \textbf{IA (\%)} & \textbf{SNS (\%)} \\
\hline
32  & 15.2 & 24.3 & 3.2 & 8.3  & 15.5 & 0.0 \\
64  & 29.8 & 34.2 & 5.8 & 8.3  & 15.5 & 0.0 \\
128 & 38.6 & 36.5 & 7.2 & 15.0 & 26.7 & 2.0 \\
256 & 49.2 & 38.7 & 9.3 & 29.3 & 42.8 & 5.7 \\
\hline
\end{tabular}
}
\end{table}

\begin{table}[t]
\centering
{\footnotesize
\setlength{\tabcolsep}{2.5pt}
\caption{Profiling results for large batch sizes (stage 2 memory pressure).}
\label{tab:large-batch}
\begin{tabular}{c|ccc|ccc}
\hline
 & \multicolumn{3}{c|}{\textbf{Overall}} & \multicolumn{3}{c}{\textbf{Stage 2}} \\
\textbf{BS} & \textbf{LS (\%)} & \textbf{L2 (\%)} & \textbf{DRAM (\%)} & \textbf{LS (\%)} & \textbf{L2 (\%)} & \textbf{DRAM (\%)} \\
\hline
256   & 30.9 & 83.7 & 22.4 & 43.1 & 71.1 & 42.2 \\
512   & 34.2 & 83.1 & 23.8 & 50.0 & 69.7 & 42.1 \\
1024  & 35.1 & 80.9 & 39.5 & 51.7 & 65.3 & 57.6 \\
2048  & 41.5 & 71.7 & 53.5 & 64.3 & 46.8 & 76.9 \\
\hline
\end{tabular}
}
\end{table}

\textbf{Small batches ($\leq 256$).} The bottleneck is insufficient parallelism. As batch size grows from 32 to 256, overall active-warps (\texttt{AW}) increase from 15.2\% to 49.2\% and issue-unit active rate (\texttt{IA}) from 24.3\% to 49.2\%. Stage 2 is the primary source of underutilization: at small batches (32--64), it achieves only $\sim$8.3\% AW and $\sim$15.5\% IA, far below the 29.3\% and 42.8\% at batch 256. Its "stall-not-selected" (\texttt{SNS}) metric is near zero at small batch, indicating there is no excessive contention. In other words, the scheduler lacks ready warps to hide stalls.

\textbf{Large batches ($\geq 256$).} The bottleneck shifts to memory hierarchy pressure. Increasing batch size from 256 to 2048 raises long-scoreboard stall (\texttt{LS}) from 30.9\% to 41.5\%, lowers L2 cache hit rate (\texttt{L2}) from 83.7\% to 71.7\%, and increases DRAM throughput (\texttt{DRAM}) from 22.4\% to 53.5\%. Stage 2 dominates this degradation: its L2 hit rate drops from 71.1\% to 46.8\%, and DRAM throughput rises from 42.2\% to 76.9\%. Overly large batches overflow the L2 cache with PBS state, intermediate buffers, and key data, forcing accesses to fall back to DRAM.

\noindent \textbf{Takeaway 3:}
\emph{PBS batch size should be carefully tuned to the optimal range to balance GPU utilization and memory hierarchy pressure.}

%% file: design.tex
\label{sec:design}

\subsection{Overview}
\label{sec:design:overview}

\begin{figure}[t]
\centering
\includegraphics[width=\linewidth]{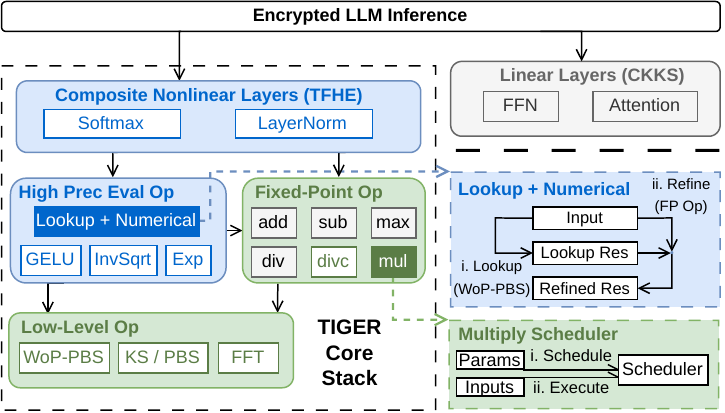}
\caption{
Architecture overview of TIGER.
TIGER targets nonlinear TFHE layers in encrypted LLM inference and organizes them into three levels: composite nonlinear layer operations, supporting operators (high-precision function evaluation and fixed-point operators), and low-level TFHE primitives.
The insets show the lookup-plus-refinement workflow and the multiplication scheduler.
Blue indicates core TIGER modules, green indicates partial design or optimization contributions, and gray denotes contextual components.
Note that GELU is an high-precision function in implementation, but also evaluated as a standalone nonlinear layer.
}
\label{fig:design:architecture_overview}
\end{figure}

Based on the above discussion, we propose \emph{TIGER}, a GPU-accelerated TFHE-based framework to enable high-performance nonlinear layers in encrypted LLM inference.
Figure~\ref{fig:design:architecture_overview} presents the overall architecture of TIGER.
At a high level, encrypted LLM inference consists of linear and nonlinear layer operations.
In our setting, linear layers such as FFN and Attention are handled in CKKS, while TIGER accelerates nonlinear layers (e.g., Softmax and LayerNorm) via TFHE.
These nonlinear layers are firstly decomposed into combinations of high-precision function evaluation and fixed-point operators, which are then further decomposed into a fine-grained sequence of low-level TFHE primitives such as WoP-PBS, KS/PBS, and FFT.
Specifically, TIGER contains the following key modules:
\begin{itemize}[leftmargin=10pt]
\setlength{\itemsep}{5pt}
    \item \textbf{Module 1: Composite Nonlinear Layers.}
    This module includes composite nonlinear layers such as Softmax and LayerNorm.
    Both Softmax and LayerNorm share a common structure of element-wise nonlinear function evaluation and fixed-point arithmetic (e.g., sum, max, division in Softmax; mean, square in LayerNorm), hence TIGER implements them in a unified way by decomposing them into a sequence of high-precision function evaluation (\emph{Module 2}) and fixed-point operators (\emph{Module 3}).

    \item \textbf{Module 2: High-precision Function Evaluation.}
    This module includes nonlinear functions such as Exp, GELU, and InvSqrt.
    It surpasses the precision limitations of existing LUT evaluation methods while keeping the computational overhead manageable.
    To this end, TIGER combines WoP-PBS-based lookup with numerical refinement.
    Specifically, the input is partitioned into multiple parts: the primary part is used to obtain an initial approximation via WoP-PBS, and the remaining parts are then processed with fixed-point operators under the guidance of first-order Taylor expansion, which further refines the result and improves evaluation precision.
    Note that GELU also serves as a standalone nonlinear layer from the view of LLM semantics, but is placed here because its implementation is directly realized by lookup-plus-refinement rather than by composition of multiple higher-level operators.

    \item \textbf{Module 3: Fixed-point Operators.}
    This module includes basic fixed-point operators such as addition, subtraction, multiplication, division, maximum, division by constants, and related carry-handling operations, most of which are built on top of standard TFHE primitives.
    In order to reduce unnecessary computation and utilize GPU parallelism more effectively, TIGER further introduces additional design and optimization for operators like multiplication and division by a constant, which are critical for achieving high performance in the context of encrypted LLM inference.
    We transform plaintext constant division into equivalent multiplication to reduce computational overhead and improve parallelism. And we use a scheduler to pre-schedule the multiplication process, in order to eliminate unnecessary computation and enhance parallelism.

    \item \textbf{Module 4: Low-level Operators.}
    This module contains the underlying TFHE building blocks used by upper layers, including WoP-PBS, KS/PBS, and FFT.
    Considering that FFT and PBS dominate the runtime of TFHE primitives, TIGER introduces dedicated low-level GPU optimizations to improve utilization and throughput.
    For FFT, we design customized kernels tailored to TFHE polynomial arithmetic, keep intermediate states in shared memory, and introduce optimizations such as 4-radix expansion and Karatsuba complex multiplication to reduce computation and synchronization cost.
    For PBS, we group PBS operations generated by upper-layer operators and apply a split-batch strategy that partitions a large batch into multiple throughput-efficient sub-batches, avoiding resource contention while preserving high GPU utilization.
    These optimizations provide an efficient execution backbone for the upper-layer operators.
\end{itemize}

\subsection{Composite Nonlinear Layer Operations}
\label{sec:design:layer-op}

\begin{algorithm}[t]
\footnotesize
\caption{Softmax Operation}
\label{alg:softmax}
\begin{algorithmic}
\Require Input vector $\mathbf{x} = (x_1, x_2, ..., x_n)$ of fixed-point numbers
\Ensure Output vector $\mathbf{y} = (y_1, y_2, ..., y_n)$ where $y_i = \frac{e^{x_i}}{\sum_{j=1}^{n} e^{x_j}}$
\State $x_{max} \gets \text{Max}(x_1, x_2, ..., x_n)$
\ForAll{$i \in [1, n]$}
    \State $z_i \gets x_{max} - x_i$
    \State $e_i \gets \text{Exp}(-z_i)$
\EndFor
\State $S \gets \sum_{i=1}^{n} e_i$
\ForAll {$i \in [1, n]$}
    \State $y_i \gets e_i / S$
\EndFor
\State \Return $\mathbf{y}$
\end{algorithmic}
\end{algorithm}

\begin{algorithm}[tb]
\footnotesize
\caption{LayerNorm Operation}
\label{alg:layernorm}
\begin{algorithmic}
\Require Input vector $\mathbf{x} = (x_1, x_2, ..., x_n)$ of fixed-point numbers, learnable parameters $\gamma, \beta$, constant $\varepsilon$
\Ensure Output vector $\mathbf{y} = \gamma \cdot \frac{\mathbf{x} - \mu}{\sqrt{\sigma^2 + \varepsilon}} + \beta$
\State $\mu \gets \sum_{i=1}^{n} x_i / n$ \Comment{Compute mean via constant division}
\ForAll{$i \in [1, n]$}
    \State $d_i \gets x_i - \mu$ \Comment{Subtract mean}
\EndFor
\State $\sigma^2 \gets \sum_{i=1}^{n} d_i^2 / n$ \Comment{Compute variance}
\State $v \gets \sigma^2 + \varepsilon$
\State $inv \gets \text{InvSqrt}(v)$ \Comment{Compute $1/\sqrt{v}$}
\ForAll{$i \in [1, n]$}
    \State $n_i \gets d_i \times inv$ \Comment{Normalize}
    \State $y_i \gets n_i \times \gamma + \beta$ \Comment{Scale and shift}
\EndFor
\State \Return $\mathbf{y}$
\end{algorithmic}
\end{algorithm}

To translate the high-level mathematical definitions of Softmax and LayerNorm into efficiently executable sequences of fixed-point operators, we briefly review the key steps of their computation and highlight the intermediate transformations required to better align them with the execution model of the underlying operators and primitives.

As described in Algorithm~\ref{alg:softmax}, \emph{Softmax} takes an input vector $\mathbf{x} = (x_1, x_2, \dots, x_n)$ and computes $y_i = \frac{e^{x_i}}{\sum_{j=1}^{n} e^{x_j}}$. In practical implementations, one typically first computes $x_{\max} = \max(x_1, \dots, x_n)$ and then replaces $e^{x_i}$ with $e^{x_i - x_{\max}}$ to avoid numerical overflow. Since $x_i - x_{\max} \leq 0$, the value of $e^{x_i - x_{\max}}$ remains well bounded.
However, in fixed-point implementations, representing $x_i - x_{\max}$ as a signed value increases implementation complexity, including bit-width handling and subsequent lookup-table access. 
As a result, our design computes $z_i = x_{\max} - x_i \geq 0$ instead, which can be directly represented as an unsigned fixed-point number, and the corresponding nonlinear function evaluation becomes the computation of $e^{-z_i}$.

As described in Algorithm~\ref{alg:layernorm}, \emph{LayerNorm} takes an input vector $\mathbf{x}$ and computes $\mathbf{y} = \frac{\mathbf{x} - \mu}{\sqrt{\sigma^2 + \varepsilon}} \cdot \gamma + \beta$, where $\mu$ is the mean, $\sigma^2$ is the variance, $\gamma$ and $\beta$ are learnable parameters, and $\varepsilon$ is a small constant introduced to avoid division by zero. For ease of implementation, we set $\varepsilon = 2^{-16}$. This choice ensures that the input to the subsequent square-root-related computation starts from a negative power of two, which simplifies the later discussion and design of interval partitioning and lookup-table inputs.

The key operation in LayerNorm is calculating the square root of the variance term and the subsequent scaling step.
A straightforward approach is to compute $S = \sqrt{\sigma^2 + \varepsilon}$ and then divide each element by $S$, which requires one lookup-table evaluation and $n$ division operations. However, division is computationally expensive and exhibits a sequential computation pattern. Instead, TIGER computes $I = \mathrm{InvSqrt}(\sigma^2 + \varepsilon)$ and then multiplies each element by $I$, which requires one lookup-table evaluation and $n$ multiplication operations. The second approach is more efficient as multiplication shows higher parallelism than division.

\subsection{High-Precision Function Evaluation}
\label{sec:design:high-prec-func-eval}

TIGER can effectively address the precision limitation of current LUT evaluation methods.
The core idea is to combine the existing lookup table estimation capabilities of WoP-PBS with numerical algorithms to further improve precision on top of the original lookup results using basic operators.

\noindent \textbf{Exp(-x).} The exponential function is fundamental to softmax computation.
As we have subtracted the maximum value from the input for numerical stability, the input to the exponential function is non-positive, which means that we only need to evaluate $\exp(x)$ in $(-\infty, 0]$, also equivalent to evaluating $\exp(-x)$ in $[0, +\infty)$.
Assuming the input is an unsigned fixed-point number with 12-bit integer part and 20-bit fractional part ($x = [x_{int}:12][x_{frac}:20] / 2^{20}$), we decompose it as $x = x_0 + x_1 + x_2$ where $x_0$ is the highest 6 bits, $x_1$ is the middle 20 bits, and $x_2$ is the lowest 6 bits.
Then $\exp(-x) = \exp(-x_0) \exp(-x_1) \exp(-x_2) = y_0 \cdot y_1 \cdot y_2$. For $y_2$ with small $x_2$, we approximate $y_2 = e^{-x_2} \approx 1 - x_2$ using the first-order Taylor expansion~\cite{bakas2024taylor, powers2015approximation}, which provides sufficient precision. For $y_1 = \exp(-x_1)$, we use WoP-PBS lookup. For $y_0$, since $\exp(-2^6)$ is extremely small, we can perform logical operations based on whether $x_0 \neq 0$: if $x_0 \neq 0$, the result is 0; otherwise, the result is $y_1 \cdot y_2$. This decomposition achieves the required precision while minimizing computational overhead.

\noindent \textbf{GELU(x).} The Gaussian Error Linear Unit (GELU) activation has been widely used in modern transformers. Assuming the input is a signed fixed-point number with 12-bit integer part and 20-bit fractional part, we first handle the sign by taking the absolute value and preserving the sign separately, focusing on $x \geq 0$. For $x \geq 16$, we observe that GELU(x) $\approx$ x within the error tolerance, so we handle this case separately. For $0 \leq x < 16$, we decompose as $x = t + \delta$ where $t$ corresponds to the lookup table input (20 bits) and $\delta$ is the small remainder (4 bits). Using first-order Taylor expansion around $t$:
\begin{align*}
\text{GELU}(x) &= (t + \delta)\Phi(t + \delta) \\
&\approx (t + \delta)(\Phi(t) + \Phi'(t)\delta) \\
&\approx \text{GELU}(t) + (\Phi(t) + t\Phi'(t))\delta \\
&= \text{GELU}(t) + G(t)\delta
\end{align*}
where $\Phi(t)$ is the standard normal CDF and $G(t) = \Phi(t) + t\Phi'(t)$. We perform a single lookup to obtain both $\text{GELU}(t)$ and $G(t)$, then compute the final result with one multiplication and one addition. Through exhaustive evaluation, we find that GELU requires $10 + 22$ bits and G requires $2 + 10$ bits of precision to satisfy the final accuracy requirements. For negative inputs, we use the identity $\text{GELU}(-x) = \text{GELU}(x) - x$.

\noindent \textbf{InvSqrt(x).} The inverse square root is needed for variance computation in LayerNorm. With 12-bit integer part and 40-bit fractional part (as we sum up all the squares of $x_i - \bar{x}$), the function characteristics vary significantly across different ranges. We adopt a piecewise lookup table strategy with small-value approximation: the input domain is divided into intervals of $[2^{-16}, 2^{-12})$, $[2^{-12}, 2^{-6})$, $[2^{-6}, 2^{2})$, $[2^{2}, 2^{12})$, with separate lookup tables for each region. Within each region, we denote $x=t+\delta$ where $t$ is the lookup table input (20 bits) and $\delta$ is the small remainder (precision varies for each region), and use Taylor expansion around $t$ to further refine the result: $1/\sqrt{x} = 1/\sqrt{t+\delta} \approx 1/\sqrt{t} - \frac{1}{2t\sqrt{t}} \delta = F(t) - G(t) \delta$, where $F(t) = 1/\sqrt{t}$ and $G(t) = \frac{1}{2t\sqrt{t}}$. We perform a single lookup to obtain both $F(t)$ and $G(t)$, then compute the final result with one multiplication and one subtraction. For each region, we custom-tailored the input range and lookup precision to meet the result precision requirement while minimizing the lookup table size and computational overhead.

\subsection{Basic FP Operations}
\label{sec:design:basic-op}

While many basic operators can be implemented using existing techniques, we provide additional designs for multiplication and constant integer division to optimize performance.

\begin{figure}[t]
\centering
\includegraphics[width=\linewidth]{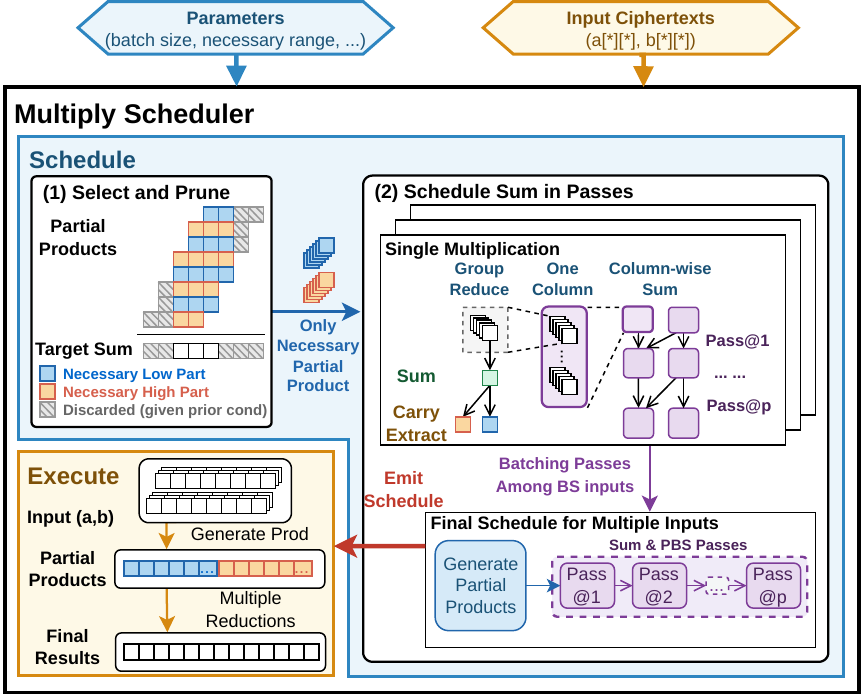}
\caption{Multiply scheduler design.
The \textbf{schedule} phase first selects only the partial products relevant to the target output range and prunes the rest.
It then organizes block-column reduction into multiple dependent passes: blocks at the same column are summed, decomposed into low parts and carries, and written back for later passes.
Passes with the same structure from different inputs can be further grouped into batched execution units.
The emitted schedule is then consumed by the \textbf{execute} phase to launch kernels and PBS operations and produce the final output blocks.}
\label{fig:design:mul-scheduler}
\end{figure}
\noindent \textbf{Multiplier Design.}
For two inputs of size $n$, multiplication produces partial products $a_i \cdot b_j, \ i,j \in [0,n-1]$, 
which must be accumulated into the corresponding output positions.
Since the reduction of these partial products involves repeated column-wise summation and carry propagation, the computation must be organized into multiple dependent passes.

To reduce unnecessary work and improve PBS utilization, we introduce a \emph{Multiply Scheduler}, as illustrated in Figure~\ref{fig:design:mul-scheduler}.
The scheduler separates multiplication into an offline \textbf{schedule} phase and an online \textbf{execute} phase.
In the schedule phase, it first identifies and prunes partial products that do not contribute to the requested output range, using prior information about the significant input region and the required output blocks.
It then plans a sequence of reduction passes, in which blocks in the same column are summed directly. The resulting sums are decomposed into low parts and carries for subsequent passes.
For independent multiplications under the same parameterization, the scheduler can further batch the same pass across inputs to improve packing efficiency and reduce fragmented PBS launches.
The resulting execution plan is emitted once and reused in the execute phase, where kernels and PBS operations are launched according to the precomputed schedule.

\noindent \textbf{Fast Division with Plaintext Integer Divisor.} If the divisor is given in plaintext, division can also be implemented as multiplication by the reciprocal instead of using a traditional division algorithm. For divisor $D$, we represent $1/D$ as an infinite base-$B$ (where $B=4$) expansion: $1/D = 0.b_0 b_1 b_2 ...$. Multiplying the dividend by each digit and summing the results yields the division result~\cite{fast-div}. This approach is more efficient than traditional division and is used for the mean computation (division by constant $n$) in LayerNorm.

\subsection{Low-Level Operator}
\label{sec:design:impl}

\noindent \textbf{FFT Implementation and Optimization.} We construct FFT as a device function where each thread block manages one polynomial's FFT to fully utilize shared memory. To handle the negacyclic property of TFHE schemes, we employ the Tangent FFT/Complex Twist technique to transform the polynomial to $\mathbb{C}[X]/(X^N + 1)$ suitable for conventional FFT. We adopt the Stockham FFT algorithm using shared memory as a ping-pong buffer. Key optimizations include: (1) 4-Radix FFT expansion that merges two passes into one, reducing synchronization overhead and multiplications; (2) Using i64 instead of f64 for $\mathbb{C}[X]$ arithmetic to reduce floating-point unit pressure; (3) Karatsuba algorithm to reduce multiplication counts in complex multiplication; (4) Special handling of the first two iterations where twiddle factors are trivial values; (5) Precomputation of twiddle factors to avoid redundant calculation.

\noindent \textbf{PBS Batching and Splitting.}
To improve GPU utilization, fixed-point operators pack multiple PBS operations into a single batch, across both intra-input and inter-input parallelism. However, as Section~\ref{sec:motiv:pbs-batch} shows, performance degrades beyond a certain batch size due to memory pressure and resource contention. Therefore, we adopt a split strategy for batch PBS execution: when receiving PBS calls, we split them into multiple small batches with sizes in the range that achieves optimal throughput (in our case, [192, 320]) and execute them in sequence. This strategy balances parallelism with resource utilization, ensuring high performance without overwhelming the GPU resources.
Though we currently determine the optimial range by offline profiling, adaptive strategies like binary search or heuristic-based algorithms could be introduced as future work to dynamically find the optimal batch size based on the specific hardware characteristics, further enhancing performance without manual tuning.

%% file: evaluation.tex
\label{sec:evaluation}

\subsection{Methodology}
\label{sec:eval:method}

\noindent \textbf{Evaluation Setup.}
We evaluate TIGER and baselines on a server equipped with NVIDIA RTX 6000 Ada Generation GPU with 48GB memory, an Intel(R) Xeon(R) Gold 5320 CPU at 2.20GHz, and 4TB of DRAM.
We implemented TIGER with over 9500 lines of CUDA C++ code, and compared it against:
\begin{itemize}[leftmargin=10pt]
\setlength{\itemsep}{5pt}
\item \emph{CPU-WoP:}
A CPU-based version of TFHE-based nonlinear layer evaluation using the same WoP-PBS algorithm, which constructs fixed-point arithmetic by referring to large-integer arithmetic in TFHE-rs~\cite{TFHE-rs} and is implemented in C++ with OpenMP.
The CPU Baseline uses 16 threads in the evaluation.
\item \emph{GPU-FBT:}
Replacing the WoP-PBS algorithm in TIGER with the FBT-TFHE algorithm, while keeping other optimizations intact, to evaluate the impact of the underlying LUT evaluation method.
\end{itemize}

\noindent \textbf{TFHE Parameters.}
Our implementation adopts the TFHE scheme with a three-level key hierarchy to support WoP-PBS. 
Table~\ref{tab:tfhe-dimensions} summarizes the LWE dimension $n$ and the TLWE polynomial degree $N$ used at each level. 
To support high-precision nonlinear computation, we instantiate the complete TFHE key hierarchy, including bootstrapping keys (BKs) and key switching keys (KSKs). 
The BKs are used for CMUX-tree evaluation, general programmable bootstrapping (GPBS), and level-2 bootstrapping, while the KSKs enable ciphertext conversion across different levels. 
Their decomposition parameters and noise standard deviations are jointly listed in Table~\ref{tab:tfhe-key-params}.

\begin{table}[]
\centering
\caption{TFHE dimension parameters.}
\label{tab:tfhe-dimensions}
\resizebox{\linewidth}{!}{%
\begin{tabular}{ccc}
\hline
Level & LWE Dimension $n$ & TLWE Polynomial Degree $N$ \\
\hline
Level 0 (CMUX/KS) & 500 & --- \\
Level 1 (GPBS) & 1024 & 1024 \\
Level 2 (Bootstrap) & 2048 & 2048 \\
\hline
\end{tabular}%
}
\end{table}

\begin{table}[]
\centering
\caption{TFHE key parameters.}
\label{tab:tfhe-key-params}
\small
\resizebox{\linewidth}{!}{%
\begin{tabular}{llcccc}
\hline
Type & \shortstack[c]{Operation\\/ Direction} & \shortstack[c]{Base Log\\/ Base Bits} & \shortstack[c]{Decomp. Length\\/ Key Length} & \shortstack[c]{Standard\\Deviation $\sigma$} \\
\hline
BK & CMUX & 6 & 3 & $2^{-15}$ \\
BK & GPBS & 4 & 6 & $7.18 \times 10^{-9}$ \\
BK & Bootstrap (Level 2) & 9 & 6 & $2^{-45}$ \\
KSK & Level 1 $\rightarrow$ Level 0 (CMUX) & 2 & 7 & $2^{-15}$ \\
KSK & Level 1 $\rightarrow$ Level 0 (GPBS) & 1 & 14 & $10^{-5}$ \\
KSK & \makecell[l]{Level 1 $\rightarrow$ Level 1\\(for FBT-TFHE)} & 6 & 3 & $2^{-25}$ \\
KSK & Level 2 $\rightarrow$ Level 1 & 2 & 16 & $2^{-31}$ \\
\hline
\end{tabular}%
}
\end{table}

\begin{table}[]
\centering
\caption{Fixed-point number format.}
\label{tab:fp-format}
\resizebox{\linewidth}{!}{%
\begin{tabular}{lcccc}
\hline
Usage & Total Bits & Integer Bits & Fractional Bits & Ciphertext Count \\
\hline
Softmax / GELU & 32 & 12 & 20 & 16 \\
LayerNorm & 34 & 14 & 20 & 17 \\
\hline
\end{tabular}%
}
\end{table}

For numerical representation, we use a radix-4 fixed-point encoding, where each LWE ciphertext carries 2 bits of information. This design provides an effective trade-off between ciphertext count and arithmetic precision. As shown in Table~\ref{tab:fp-format}, we use a 32-bit fixed-point format for general computation and a slightly wider 34-bit format for LayerNorm to accommodate its larger dynamic range.

\subsection{Overall Performance}
\label{sec:eval:overall}

\begin{figure}[]
\vspace{-5pt}
\centering
  \includegraphics[width=0.9\linewidth]{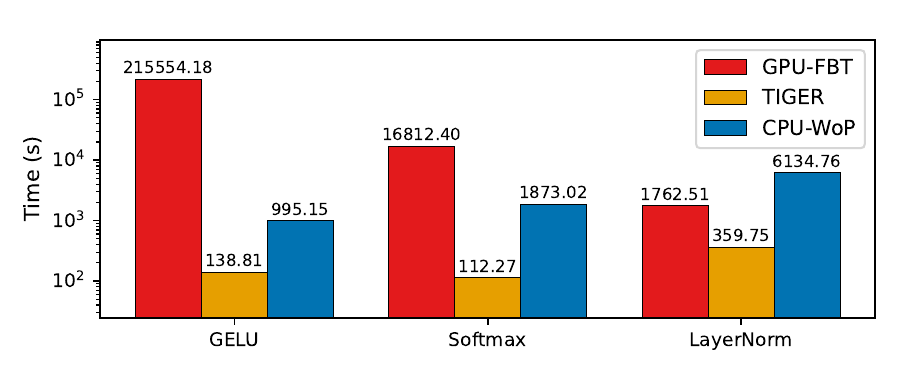}
  \vspace{-15pt}
\caption{Layer-wise execution time comparison among TIGER, CPU-WoP, and GPU-FBT for GELU, Softmax, and LayerNorm (log scale).}
\label{fig:layer-time}
\end{figure}

\noindent \textbf{Layer-wise Performance Comparison.}
We first evaluate the execution time of each nonlinear operator, including GELU, Softmax, and LayerNorm, and compare TIGER against the baseline implementations.
As shown in Figure~\ref{fig:layer-time}, TIGER consistently achieves the lowest latency across all three operators.
Compared with the CPU baseline, TIGER delivers 7.17$\times$, 16.68$\times$, and 17.05$\times$ speedups for GELU, Softmax, and LayerNorm, respectively.
TIGER also consistently outperforms GPU-FBT, highlighting the efficiency advantage of its WoP-PBS-based design over the FBT-TFHE alternative.

The performance improvement stems from multiple factors: (1) the massive parallelism available on GPU for processing multiple ciphertexts simultaneously, (2) the batched PBS operation with split strategy that maximizes GPU utilization, and (3) the optimized FFT implementation using 4-radix decomposition and Karatsuba multiplication.
The speedup is more pronounced for Softmax and LayerNorm, whose computation involves more complex combination of different fixed-point operators and therefore benefits more substantially from GPU parallelism.
Overall, these results show that TIGER is particularly effective for nonlinear layers that dominate the latency of privacy-preserving Transformer inference.

\noindent \textbf{End-to-End Inference Performance.}
\label{sec:eval:e2e-perf}
To translate TIGER's layer-wise gains into a realistic deployment setting,
we estimate the end-to-end latency of a single GPT-2 Transformer block,
which comprises one CKKS-based linear layer, four CKKS$\leftrightarrow$TFHE
protocol conversion steps, and the four nonlinear operators evaluated in this work
(2$\times$LayerNorm, 1$\times$Softmax, 1$\times$GELU).
The CKKS components are implemented in OpenFHE with ring dimension $N=2^{16}$ and batched packing (128/768 active slots, padded to 128/1024), with conversion performed at the corresponding batch granularity.
Table~\ref{tab:e2e-perf} summarizes the per-component latency breakdown.

\begin{table}[]
\centering
\caption{Estimated single Transformer block latency breakdown (s).}
\label{tab:e2e-perf}
\begin{tabular}{lccr}
\hline
\textbf{Component} & \textbf{TIGER} & \textbf{CPU Baseline} & \textbf{Speedup} \\
\hline
Linear (CKKS)              & 0.29  & 26.21   &  \\
CKKS$\rightarrow$TFHE (4$\times$) & 0.20  & 8.95    &  \\
TFHE$\rightarrow$CKKS (4$\times$) & 14.60 & 148.77  &  \\
Nonlinear (TFHE)           & 970.58 & 15137.68 & 15.60$\times$ \\
\hline
\textbf{Total}             & \textbf{985.66} & \textbf{15321.62} & \textbf{15.54$\times$} \\
\hline
\end{tabular}
\end{table}

As shown in Table~\ref{tab:e2e-perf}, TIGER reduces the estimated latency of a full Transformer block from 15321.62~s to 985.66~s, achieving an overall speedup of \textbf{15.54$\times$}.
This end-to-end gain is nearly identical to the 15.60$\times$ speedup of the nonlinear component, indicating that block-level latency is dominated by TFHE-based nonlinear evaluation.

The breakdown in Table~\ref{tab:e2e-perf} confirms this trend: nonlinear operators account for 970.58~s of the 985.66~s total runtime on TIGER, i.e., approximately 98.5\%.
By contrast, TFHE$\rightarrow$CKKS conversion contributes about 1.5\%, while the CKKS linear layer and CKKS$\rightarrow$TFHE conversion are negligible.
Since the linear CKKS path and protocol conversion have already been effectively optimized in prior work, their estimated contribution to end-to-end latency is relatively small, leaving nonlinear TFHE computation as the dominant bottleneck targeted by TIGER.

Overall, these results show that accelerating nonlinear TFHE operators is the key to reducing end-to-end inference latency in privacy-preserving Transformer inference.

\subsection{Deep Dive into TIGER's Performance}
\label{sec:eval:deep-dive}

\begin{figure*}[]
\centering
\includegraphics[width=\linewidth]{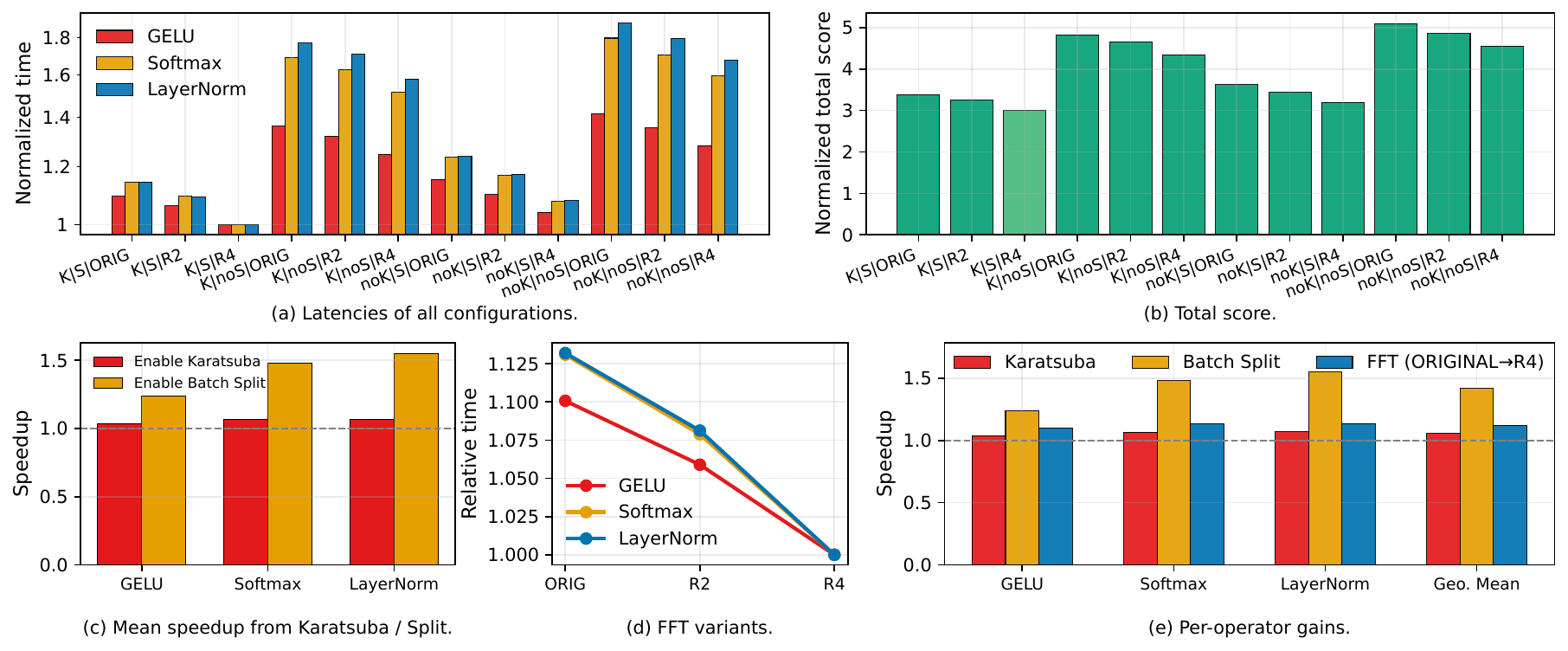}
\caption{Ablation study.}
\label{fig:ablation-study}
\end{figure*}

\noindent \textbf{Ablation Study of Low-level Optimizations.}
To understand the contribution of each optimization in TIGER, we conduct an ablation study over three optimization axes: (1) Karatsuba multiplication for complex FFT operations (K/noK), (2) KSPBS batch splitting strategy (S/noS), and (3) FFT implementation variants, including ORIG (original unoptimized version), R2 (optimized radix-2), and R4 (optimized radix-4). In total, this yields 12 configurations, which we evaluate on all three nonlinear operators.

The best configuration, which enables Karatsuba, Split, and R4 FFT simultaneously, consistently achieves the lowest runtime among all tested settings. Compared to the baseline with all optimizations disabled, it reduces execution time from 31.29 s to 22.06 s for GELU, from 202.41 s to 112.64 s for Softmax, and from 119.41 s to 63.39 s for LayerNorm, corresponding to 1.42$\times$, 1.80$\times$, and 1.88$\times$ speedups, respectively.

Figure~\ref{fig:ablation-study} summarizes the ablation results from five complementary views. Subfigures (a) and (b) provide a configuration-level view by showing the normalized latency of all 12 configurations and their aggregated total scores, respectively. Subfigures (c) and (d) isolate the contribution of individual optimization axes by summarizing the gains from Karatsuba and Batch Split and by comparing FFT variants. Finally, subfigure (e) provides a compact per-operator summary of the gain brought by each optimization.

The ablation study yields three main findings:
\begin{enumerate}
    \item \textbf{KSPBS Split Strategy.} Enabling the split strategy provides the largest performance gain, typically reducing execution time by about 25--30\%. This effect is most visible in subfigures (c) and (e), and comes from avoiding cache thrashing by keeping the batch size within the optimal range [192, 320].

    \item \textbf{Karatsuba Multiplication.} Karatsuba multiplication consistently improves performance across all three operators, with gains of roughly 5--10\%, as shown in subfigures (c) and (e). The improvement comes from reducing the number of floating-point multiplications in complex FFT operations.

    \item \textbf{4-Radix FFT.} Moving from ORIG or R2 to R4 further improves runtime by about 5--8\%, as illustrated in subfigure (d). This optimization effectively merges two FFT passes into one, reducing synchronization overhead and the total number of multiplications.
\end{enumerate}

Taken together, subfigures (c), (d), and (e) show that the KSPBS split strategy contributes the most to overall performance, followed by the 4-radix FFT optimization and Karatsuba multiplication.

\begin{figure*}[]
\centering
\includegraphics[width=\linewidth]{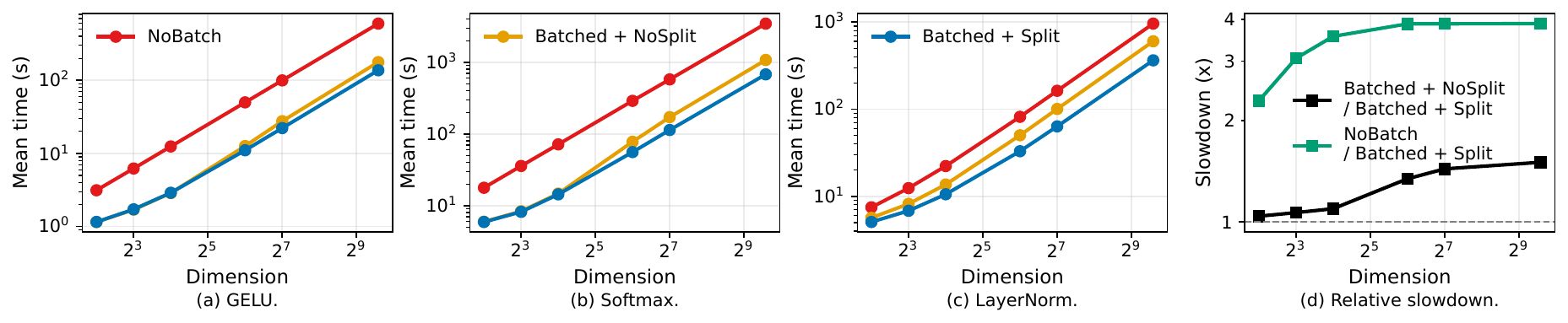}
\caption{Scalability analysis.}
\label{fig:scalability-log}
\end{figure*}

\noindent \textbf{Scalability Analysis.}
We evaluate the scalability of TIGER across different input dimensions (4, 8, 16, 64, 128, and 768), which correspond to various LLM hidden sizes. We test three configurations:
\noindent \textbf{NoBatch} applies no batching optimization, \textbf{Batched + NoSplit} enables batching without the split strategy, and \textbf{Batched + Split} uses the full optimization with both batching and split strategy.

Figure~\ref{fig:scalability-log} presents the scalability results on a logarithmic scale: subfigures (a)--(c) show the execution time trends of GELU, Softmax, and LayerNorm, respectively, while subfigure (d) summarizes the relative slowdown of Batched + NoSplit and NoBatch against Batched + Split.

\begin{enumerate}
    \item At small dimensions (4, 8), the performance difference between Batched + NoSplit and Batched + Split modes is minimal because the batch size naturally falls within the optimal range.

    \item At larger dimensions (128, 768), the Batched + NoSplit mode suffers from cache thrashing due to excessively large batches, while the Batched + Split mode with split strategy maintains optimal performance by keeping each sub-batch within [192, 320] range.

    \item The performance ratio between NoBatch and Batched + Split modes increases with dimension, reaching approximately 4.35$\times$ for GELU at dimension 768.
\end{enumerate}

These results validate that TIGER's batch and split strategy is essential for achieving optimal GPU utilization, especially when processing large-scale inputs typical of real LLM inference workloads.

%% file: discussion.tex
\label{sec:discussion}

\noindent \textbf{Multi-GPU Scaling.}
In TIGER, high-precision function evaluation and most fixed-point operators of ciphertext groups are element-wise and independent.
As a result, inputs from the same nonlinear layer can be partitioned across devices with little cross-GPU dependency.
The main exception is reduction-style operators such as sum and max. Nevertheless, they are invoked infrequently and not computationally costly, so the inter-device communication is limited.

From the memory perspective, multi-GPU execution is also practical.
In our implementation, secret keys and LUT occupy only several GBs and remain immutable after initialization.
Thus, each GPU can hold a local replica of them, while ciphertext inputs, outputs, and intermediate buffers are sharded across devices. Given that modern data-center GPUs typically provide tens of GBs of memory, such replication is affordable and avoids frequent remote accesses to shared key material.
These properties suggest that TIGER can easily scale to multiple GPUs by leveraging an owner-computes partitioning principle under a bulk-synchronous parallel execution model.


\noindent \textbf{Portability.}
Although TIGER is developed for encrypted LLM inference, its core design is also applicable to many privacy-preserving inference workloads that require high-precision nonlinear computation.
Representative directions include encrypted probabilistic prediction and risk scoring for clinical or financial assessment~\cite{hosmer2013applied,steyerberg2019clinical,maldonado2022deepcredit}, as well as privacy-preserving scientific or biomedical analysis with continuous-valued outputs~\cite{bishop2006prml,rasmussen2006gaussian,karniadakis2021physics,miotto2018deep}.
In such settings, approximation errors in sigmoid-, softmax-, or exponentiation-related computation may directly affect confidence estimation, threshold-based decisions, risk ranking, or the fidelity of continuous-valued predictions~\cite{niculescu2005predicting,guo2017calibration}.
Compared with LLM inference, where moderate approximation errors in nonlinear layers may sometimes be tolerated if the final token distribution is largely preserved, these applications are often more sensitive to the precision of nonlinear evaluation.
Therefore, beyond transformers, TIGER can also serve as a useful TFHE substrate for encrypted inference tasks whose outputs intrinsically depend on accurate nonlinear computation.

%% file: conclusion.tex
\label{sec:conclusion}
This paper presents \emph{TIGER}, the first GPU-accelerated framework for high-precision TFHE-based nonlinear layer evaluation in encrypted LLM inference.
By combining GPU-efficient WoP-PBS, numerical refinement, and batch-oriented execution, TIGER bridges the gap between low-level TFHE primitives and practical model-level nonlinear operators.
Our implementation supports nonlinear layers used in GPT-2, including GELU, Softmax, and LayerNorm, while significantly outperforming a CPU baseline.
We hope TIGER can serve as a practical foundation for future high-precision encrypted inference systems on GPUs.